\documentclass{article}

\usepackage{epsfig}
\usepackage{subfigure}
\begin{document}

\title{A model for hierarchical patterns under mechanical stresses}
\author{F. Corson$^{(a)}$, H. Henry $^{(b)}$, M. Adda-Bedia$^{(a)}$\\
$^{(a)}$ Laboratoire de Physique Statistique, \'Ecole Normale
Sup\'erieure, CNRS,\\
24 rue Lhomond, 75231 Paris Cedex 05, France\\
$^{(b)}$ Physique de la Mati\`ere Condens\'ee,
\'Ecole Polytechnique, CNRS,\\ rte de Saclay, 91128 Palaiseau, France. }
\maketitle

\begin{abstract}
We present a model for mechanically-induced pattern formation in growing biological tissues
and discuss its application to the development of leaf venation networks.
Drawing an analogy with phase transitions in solids,
we use a phase field method to describe the transition between two states of the tissue,
e.g. the differentiation of leaf veins,
and consider a layered system where mechanical stresses are generated by differential growth.
We present analytical and numerical results for one-dimensional systems,
showing that a combination of growth and irreversibility gives rise to hierarchical patterns.
Two-dimensional simulations suggest that such a mechanism could account for
the hierarchical, reticulate structure of leaf venation networks,
yet point to the need for a more detailed treatment of the coupling
between growth and mechanical stresses.
\end{abstract}

\section{Introduction}

A broad range of patterns in nature are induced by mechanical forces.
While the most well understood examples are found in physical phenomena,
e.g. in solids, fracture~\cite{freund90},
wrinkling~\cite{cerda-03}, delamination~\cite{audoly-99},
or localized deformation~\cite{gioia-01},
it is increasingly becoming appreciated that mechanical forces
are also essential in the morphogenesis of living organisms.
In the simplest case, biological tissues can undergo mechanical instabilities
as a result of inhomogeneous growth~\cite{benamar-05,sharon-07}.
More broadly, the importance of the mechanical and adhesion properties
of cells in morphogenesis~\cite{lecuit-07}
and the ability of mechanical forces to act as a signal in growing tissues and organisms
~\cite{farge-03,nelson-05,hufnagel-07}
suggest that the interplay between mechanical forces and biochemical factors
has a central role in development.

An illustration of these ideas can be found in
recent investigations of the structure of leaf venation networks~\cite{couder-02}.
The standard model for vein formation
involves the transport of the plant hormone auxin~\cite{sachs-81}:
auxin circulates through the leaf, with the leaf base acting as a sink;
as the leaf develops, auxin flow becomes progressively
concentrated along preferential paths, which mark the location of future veins.
This model has received ample experimental support
(see e.g.~\cite{scarpella-06}),
yet it fails to explain certain aspects of leaf venation,
such as the presence of closed loops.
Simulations of auxin transport generally
give rise to tree-like structures~\cite{feugier-05},
a typical property  of diffusion-limited growth processes~\cite{branching}.
In contrast,
it has been noted that the reticulate structure of venation networks
is very similar to that of crack patterns,
suggesting the possibility that the differentiation of leaf veins
could be governed by mechanical stresses~\cite{couder-02}.
The inner tissues of the leaf, where the veins form,
are subjected to compressive stresses by the epidermis,
and one of the earliest signs of vein differentiation is the elongation of vascular cells~\cite{nelson-97},
which could release the compressive stresses in their vicinity,
providing a feedback between differentiation
and the mechanical state of the tissue.

In this article, we develop a theoretical approach to study
the formation of patterns induced by mechanical stresses in biological tissues.
Of particular interest is the fact that these tissues are growing as the patterns form.
We show how growth can be the driving force that causes
the formation of new structures,
and how the history of their formation
can be reflected in their organization.
Let us emphasize that this study mainly aims at demonstrating possible effects of mechanical stresses,
and that the approach developed here would have to be integrated
with biochemical mechanisms in a more complete model.

Stated in general terms, the patterning mechanism we consider is the following:
compressive stresses imparted on a tissue cause it to switch to
a ``collapsed'' state occupying a smaller volume,
whereby the compression is released.
In the case of leaf venation,
this transition describes the differentiation of veins,
and the collapsed state refers to the elongated vascular cells.
An analogy can be made with the behavior of solid foams,
which exhibit localized deformation upon compression~\cite{gioia-01}.
Our model is not explicitly constructed according to this analogy.
Instead,
we consider the initial and collapsed states of the tissue
as two phases of a continuous medium,
which we describe using a phase field method.
However,
the response of this two-phase medium
is analogous to that of a non-linear elastic material,
and can be analyzed with reference to the behavior of such materials
\footnote{Conversely, localized deformation
in non-linear elastic materials can be interpreted as a first-order phase transition.}.

In what follows, we first present our model in a one-dimensional setting.
We define a minimal model, amenable to analytical treatment,
that implements the patterning mechanism described above.
We consider a layered system,
in which the tissue is coupled to a rigid substrate,
which could represent a stiffer tissue such as the epidermis of plant leaves.
Stresses are induced by a mismatch between the two layers,
which can occur as a result of differential growth.
If the transition between the two states of the tissue is reversible,
regular patterns are obtained,
in which domains of collapsed and uncollapsed tissue alternate.
If, instead,
the transition to the collapsed state is made irreversible to model tissue differentiation,
hierarchical patterns are obtained.

When turning to two-dimensional systems,
the tensorial nature of mechanical fields comes into play,
and a much greater variety of behaviors are possible.
Rather than to attempt a systematic exploration,
we present a few illustrative examples of the patterns that can be obtained.
In particular,
we show that anisotropic collapse,
which could describe the elongation of vascular cells in plant leaves,
produces reticulate patterns comprised of interconnected stripes of collapsed tissue.
Our simulation results suggest that hierarchical, reticulate patterns
might be obtained when irreversibility is introduced.
However, in the simple model presented here,
these patterns are disrupted by strong residual stresses
that develop in the vicinity of growing collapsed regions.
We suggest that a more satisfactory model would require
a more detailed description of the interplay between mechanical stresses and growth.
We also contrast our model with a similar approach presented in~\cite{jagla}.

\section{One-dimensional model}

In this section, we present our modeling approach in the context of a one-dimensional system.
We consider a tissue that can exist in two states,
an ``initial'' state and a ``collapsed'' state that has a smaller volume.
Transition to the collapsed state is triggered by compressive stresses,
and releases the compression.
While, in general, the two states of the tissue could differ in their stiffness
(the collapse could be due to the transition to a softer state),
we assume that they only differ in their rest configurations
(the collapsed state has a smaller volume at rest).
In the case of leaf veins,
the initial and collapsed states of the tissue represent
the undifferentiated and vascular tissues, respectively,
and the change in rest configuration corresponds to
the elongation of vascular cells as they differentiate.
The tissue is described as a continuous medium,
and its two states as two ``phases'', using a phase field model.

The phase field approach, used here in the context of mechanically induced
pattern formation~\cite{jagla}, was first introduced to describe the growth
of a solid in a liquid~\cite{ColLev85,KarRap98,fix83,langer86} (reviewed in~\cite{boettinger2002,cinca2004})
and has later proved to be a powerful
tool to describe free boundary problems in fluid flows~\cite{Foletal99,BibMis03}
and solid-solid phase transitions in alloys~\cite{khachaturyan93,lebouar98}. This is especially
the case in 3D, where interface tracking methods are extremely difficult to
implement~\cite{Pusztai2005,kobayashi2005}.
Our model is similar to those developed for
fracture in~\cite{kkl,henry-08,henry-04,Spatschek2006,Pilipenko2007} and for
solid-solid phase transitions in~\cite{khachaturyan93,lebouar98},
which comprise phases having different mechanical properties.

A continuously varying auxiliary field $\phi$ is introduced to indicate the local state of the tissue,
with the convention that $\phi=0$ corresponds to the initial state,
and $\phi=1$ to collapsed regions.
The behavior of the system, i.e. both its deformations and the transitions between the two states,
is assumed to be governed by an energy functional $E\{\phi,u\}$,
where $u$ is the displacement field.
While this is natural for a physical system, it may seem rather arbitrary for a biological one.
However, we will see that the resulting behavior is sufficiently general.
The evolution of the system is assumed to be quasi-static,
i.e. at any given time it is in a state of equilibrium that minimizes $E$,
and satisfies the equilibrium equations
\begin{eqnarray}
\frac{\delta E}{\delta u}&=&0
\label{eq:mech_eq}\\
\frac{\delta E}{\delta\phi}&=&0.
\end{eqnarray}
For a one-dimensional system, the energy $E$ is defined by
\begin{equation}
\label{eq_E}
E=\int\left[\frac{D}{2}(\partial_x\phi)^2+e(\phi,\epsilon)+f(\phi)\right]dx.
\end{equation}
The first term in the integral is a squared gradient term, which penalizes sharp variations in the phase field.
$e$ is the density of elastic energy,
which depends on the phase and on the strain $\epsilon=\partial_x u$.
$f(\phi)$ is a potential that,
together with the coupling between $\phi$ and $\epsilon$ through $e$,
governs the behavior of the phase field.
Assuming that both phases (treated as linear elastic) have the same elastic modulus $\mu$
and differ only in their rest configuration,
the energy defined by eq.~\ref{eq_E} takes the form
\begin{equation}
E=\int\left[\frac{D}{2}(\partial_x\phi)^2+\frac{\mu}{2}\left(\epsilon-\epsilon_r(\phi)\right)^2+f(\phi)\right]dx,
\label{eq_E_epsr}
\end{equation}
where the rest configuration is given by
\begin{equation}
\epsilon_r(\phi)=h(\phi)\epsilon_{r1}.
\end{equation}
The function $h$ (defined below) satisfies $h(0)=0$ and $h(1)=1$,
and $\epsilon_{r1}$ is the rest configuration of the collapsed phase.
The potential $f$ is defined by
\begin{equation}
f(\phi)=\alpha h(\phi)^2+\beta h(\phi).
\end{equation}
$h(\phi)=3\phi^2-2\phi^3$
is chosen such that $h'(0)=h'(1)=0$,
so that uniform phases ($\phi=0$ or $\phi=1$) are stationary states.

Qualitatively,
it can seen be from eq.~\ref{eq_E_epsr} that under sufficient compression,
the system will tend to switch to the collapsed state to lower its energy.
The response of the system can be described by examining
its uniform states of equilibrium (where $\phi$ and $\epsilon$ are uniform).
In the range of parameters considered here
\footnote{The parameters are chosen such that the critical strains defined below satisfy $\epsilon_{c1}<\epsilon_{c0}$.},
$\phi$ has a unique equilibrium value for every value of the strain $\epsilon$.
Defining the critical strains
\begin{eqnarray}
\epsilon_{c0}&=&\frac{\beta}{\mu\epsilon_{r1}},\\
\epsilon_{c1}&=&\frac{\mu\epsilon_{r1}^2+2\alpha+\beta}{\mu\epsilon_{r1}},
\end{eqnarray}
$\phi$ is equal to 0 when $\epsilon>\epsilon_{c0}$
(note that since we are considering compressive strains,
$\epsilon$ and $\sigma$ are generally negative),
varies gradually from 0 to 1
as $\epsilon$ goes from $\epsilon_{c0}$ to $\epsilon_{c1}$,
and is equal to 1 when $\epsilon<\epsilon_{c1}$
(for strong compressive strains).
Since $\phi$ is uniquely determined by the strain $\epsilon$,
so is the stress $\sigma=\mu(\epsilon-\epsilon_r(\phi))$.
Our two-phase model thus behaves as a non-linear elastic material
with a certain stress-strain response.

\begin{figure}
\begin{center}
\subfigure[]{\includegraphics[scale=1.2]{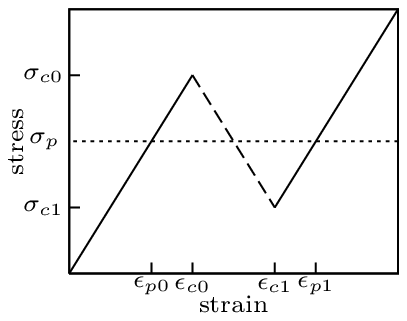}}
\subfigure[]{\includegraphics[scale=1.2]{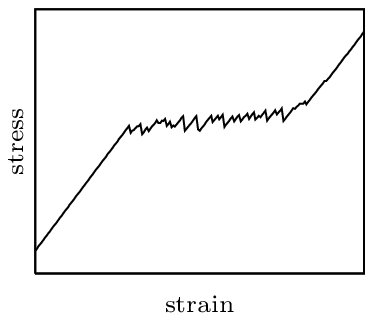}}
\end{center}
\caption{
(a) Sketch of the stress-strain curve of the model.
The dashed segment corresponds to unstable states, which phase-separate.
The dotted horizontal line indicates the plateau stress observed when the two phases coexist.
(b) Stress-strain response of a simulated inhomogeneous system.
\label{stress-strain}}
\end{figure}

As shown on fig.~\ref{stress-strain}(a), this response is non-monotonic,
and the phase transitions in our model are equivalent to
mechanical instabilities in an elastic material with a non-monotonic stress-strain curve~\cite{ericksen-75}.
The states for which the slope is negative,
which correspond to values of the phase intermediate between 0 and 1,
are unstable.
If the system is subjected to increasing strain,
it deforms uniformly until $\epsilon=\epsilon_{c0}$,
then phase-separates.
When the two phases coexist, the stress has a fixed ``plateau'' value $\sigma_p$,
as do the strains in the two phases (see fig.~\ref{stress-strain}(a)).
In a biological context, the plateau stress can be understood as a ``homeostatic stress''
restored by the formation of a collapsed region.
If the strain is further increased,
the proportion of the collapsed phase increases until the system is entirely collapsed,
after which the strain is uniform again.
Note that states intermediate between the plateau stress and the instability threshold are metastable,
and in actual physical systems, the distinction between the two can be blurred by disorder~\cite{gioia-01}.
This can be reproduced in our model
by adding a random prefactor $r(x)$ to the terms $e(\phi,\epsilon)+f(\phi)$
in the expression of the energy (eq.~\ref{eq_E}).
This creates local weak spots, so that the collapsed phase nucleates before the critical stress
for an ideal, homogeneous system (compare figs.~\ref{stress-strain}(a) and (b)).

As illustrated by the simple, piecewise linear stress-strain curve shown in fig.~\ref{stress-strain}(a),
our model provides a minimal description of a mechanism
for the release of mechanical stresses by localized deformation.
There are just enough parameters to allow independent adjustment
of the properties that will determine its pattern-forming behavior,
i.e. the instability threshold, the amount of deformation associated with the collapse,
and the equilibrium stress at the interface between the two phases.

\begin{figure}
\begin{center}
\includegraphics[scale=1.2]{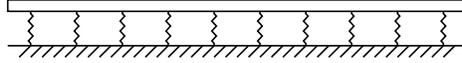}
\end{center}
\caption{Diagram of the system. A layer of tissue (top) is coupled to a rigid substrate (bottom).
\label{system}}
\end{figure}

In what follows, we consider the system shown in fig.~\ref{system},
in which a layer of tissue described by the above model is coupled to a rigid substrate.
If there is any mismatch between the two layers,
mechanical stresses are induced in the tissue.
Such a layered arrangement is commonly used to model various mechanical instabilities,
e.g. the fragmentation of thin films~\cite{meakin-87,blumen-93},
and can be likened with the layered structure of plant leaves~\cite{couder-02},
with the rigid substrate representing the epidermis.

The coupling between the two layers is described by a quadratic potential,
which included into the energy (eq.~\ref{eq_E_epsr}) yields
\begin{equation}
E=\int \left[\frac{D}{2}(\partial_x\phi)^2+\frac{\mu}{2}\left(\epsilon-\epsilon_r(\phi)\right)^2
+f(\phi)+\frac{k}{2}(u-\eta x)^2\right] dx.
\label{eq_E_eta}
\end{equation}
The parameter $k$ represents the strength of the coupling,
and $\eta$ is a measure of the mismatch between the two layers,
defined by $L_s=(1+\eta)L$,
where the $L_s$ and $L$ are the lengths of the substrate and of the tissue layer in an unstressed state.

To understand how the phase transition can be triggered,
it is useful to compute the strain field when the tissue is in its initial state ($\phi=0$)
\footnote{
Similar calculations could be carried out for
the states of equilibrium in which the two phases coexist
if the width of the interfaces is neglected.}.
The equation of mechanical equilibrium (eq.~\ref{eq:mech_eq}) yields
\begin{equation}
\mu\partial_{xx}u-k (u-\eta x)=0,
\label{equation_equlibre_mec_hh}
\end{equation}
which is linear in $u$ and admits solutions of the form
\begin{equation}
u=\eta x+Ae^\frac{x}{\lambda}+Be^{-\frac{x}{\lambda}}.
\label{eq_u}
\end{equation}
In this equation, $\lambda=\sqrt{\mu/k}$ is a characteristic elastic length scale of the system,
and $A$ and $B$ are constants that are determined by the boundary conditions.
Here, the edges of the tissue are free of stresses,
and $\partial_x u=0$ at $x=\pm L_s$.
Computing $A$ and $B$ and deriving eq.~\ref{eq_u} with respect to $x$,
we find that the strain in the tissue is given by
\begin{equation}
\epsilon=\eta\left(1-\frac{\cosh(x/\lambda)}{\cosh(L_s/\lambda)}\right).
\end{equation}
If the system is large compared to the elastic length scale $\lambda$,
the strain is close to the strain imposed by the substrate everywhere except
at the edges, and the stresses exhibit a broad plateau.
If the system size is comparable to or smaller than the elastic length scale,
the stresses have a more localized maximum at the center of the system,
and the value of this maximum  increases with the size of the system.

One way instabilities can be induced is by increasing the mismatch between the two layers
(this is analogous to the situation of a shrinking film on a substrate).
From the above,
it appears that instabilities can also be induced by growth
with the mismatch between the two layers being fixed,
that is as result of uniform growth.
The latter regime will always be considered when simulating
patterning in growing tissues\footnote{
If the two layers were instead assumed to grow at different rates,
then the mismatch would become arbitrarily large with time.
}.

\begin{figure}
\begin{center}
\includegraphics[scale=1.2]{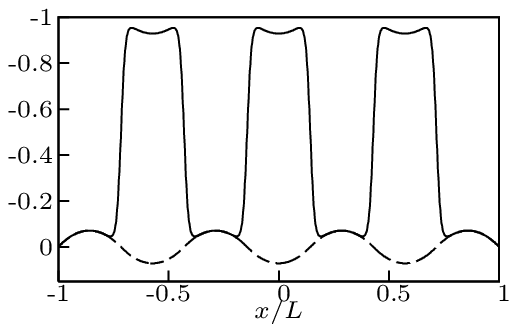}
\end{center}
\caption{
A simulated pattern with three collapsed zones.
The solid line and the dashed line represent the strain and the stress, respectively.
The parameters of the model are $\mu=1$, $\epsilon_{r1}=-1$,
$\alpha=-0.25$, $\beta=0.25$, $D=4\times 10^{-4}$, $k=16$, $L=1$, and $\eta=-.5$.
With these values, the elastic length scale is $\lambda=.25$,
the instability threshold is $\epsilon_{c0}=-0.25$ and the plateau stress is $\sigma_p=0$.
Accordingly the strains in the two phases when they coexist
are equal to the rest strains in their rest configurations,
i.e. $\epsilon_{p0}=0$ and $\epsilon_{p1}=-1$.
\label{curves}}
\end{figure}

We now turn to a numerical analysis of the evolution of the system.
So far, we specified only its energy, which determines its states of equilibrium and their stability.
To determine the new state of equilibrium reached after an instability occurs,
its dynamics must be specified.
A reasonable assumption is that mechanical relaxation
is much faster than the transition between the two states of the tissue,
so that at any time the system is at mechanical equilibrium:
\begin{equation}
0=\frac{\delta E}{\delta u}
=-\mu\partial_x\left(\partial_x u-\epsilon_r(\phi)\right)
+k(u-\eta x)
\label{eq_equilibre_mec}
\end{equation}
Dynamics need thus be specified only for the phase field.
We assume the latter obeys a relaxation equation of the form
\begin{equation}
\partial_t\phi\sim-\frac{\delta E}{\delta \phi}=D\Delta\phi-\partial_\phi e-\partial_\phi f.
\label{equation_relax}
\end{equation}
At each time step of the overall evolution of the system,
i.e. of its growth,
this relaxation equation is solved assuming the system is at mechanical
equilibrium, until an equilibrium is reached
\footnote{
The equations of mechanical equilibrium (a set of linear equations)
are solved at each step of the relaxation.
The relaxation is stopped when the rate of change of the phase field falls below a certain threshold.
}. And this state of equilibrium satisfies the equations\footnote{We are thus assuming that there are three well separated time scales
$\tau_\mathrm{mech}\ll\tau_\mathrm{diff}\ll\tau_\mathrm{growth}$,
where 
$\tau_\mathrm{mech}$, $\tau_\mathrm{diff}$, and $\tau_\mathrm{growth}$
are the characteristic time scales corresponding to
mechanical relaxation, cell differentiation, and growth, respectively.}:
\begin{equation}
\left\{
\begin{array}{ccl}
0&=&\mu\partial_x\left(\partial_x u-\epsilon_r(\phi)\right)
-k(u-\eta x)
\\
0&=&D\Delta\phi-\partial_\phi e-\partial_\phi f.
\end{array}
\right.
 \end{equation}

Most of the time, no instability occurs,
and the new state of equilibrium is very close to the previous one.
If an instability occurs,
the phase field evolves to a new state
that contains a different number of collapsed regions.
Fig.~\ref{curves} shows a typical
equilibrium configuration. 

\begin{figure}
\begin{center}
\subfigure[]{\includegraphics[scale=1.2]{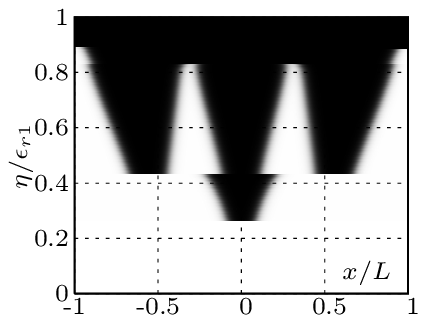}}
\subfigure[]{\includegraphics[scale=1.2]{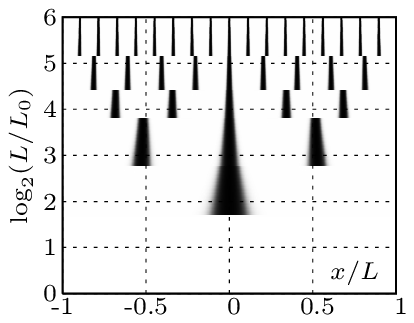}}
\\
\subfigure[]{\includegraphics[scale=1.2]{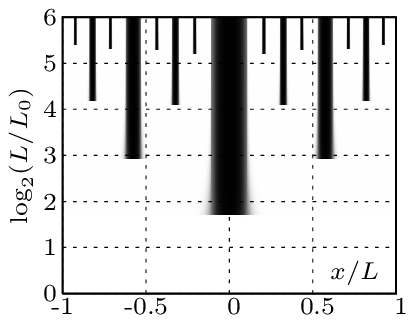}}
\subfigure[]{\includegraphics[scale=1.2]{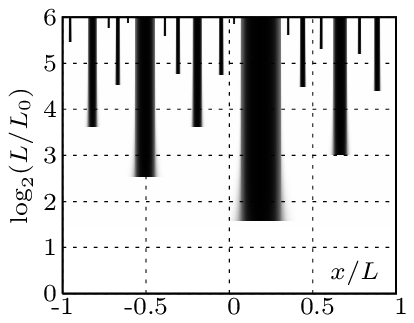}}
\end{center}
\caption{
Patterns obtained in a one-dimensional system.
These plots show the evolution of the phase field, with the collapsed regions appearing in black.
Model parameters are given in  fig.~(\ref{curves}).
(a) Increasing mismatch between the two layers,
with the size of the tissue being fixed ($L=1$).
(b-d) Growing system with a fixed mismatch ($\eta=-0.275$).
The initial size of the tissue is $L_0=0.25$.
(b) Reversible evolution. Note that the collapsed regions appear to
shrink only because the total size of the system increases.
(c) Irreversible evolution.
(d) Same as (c) with disorder.
\label{images}}
\end{figure}

The results obtained in various conditions are shown in fig.~\ref{images}.
In the first example (fig.~\ref{images}(a)),
the tissue has a fixed size and the mismatch between the two layers is progressively increased.
A limited number of collapsed regions are formed and grow to occupy the entire tissue.
In the second example (fig.~\ref{images}(b)),
the system grows with the mismatch remaining constant
($L=e^\frac{t}{\tau}L_0$ and $L_s=e^\frac{t}{\tau}L_{s0}$).
In that case, instabilities occur with time in an approximately self-similar cascade.
Note that in both of these examples,
the existing collapsed zones readjust after each new instability,
and the system relaxes to a regular pattern that minimizes its energy.
In the process, existing collapsed zones may change size or position,
which is possible because the transition between the phases is reversible.
In the case of a growing system (fig.~\ref{images}(b)),
this means that the pattern does not retain any trace of the history of its formation.

Now, if the transition to the collapsed state represents tissue differentiation,
it should be irreversible, precluding the above rearrangements.
A simple way of implementing irreversibility is to make the
evolution of the phase field $\phi$ irreversible.
Assuming that the transition must progress to a certain point before it becomes irreversible,
we allow $\phi$ to increase or decrease when it is smaller than a certain threshold ($1/2$),
and only to increase when it is larger
\footnote{
Such an \textit{ad hoc} rule would clearly be unsatisfactory to describe a physical system,
but in the case of biological systems,
it can be seen as a necessary simplification of the complex phenomena involved.}.
As shown in fig.~\ref{images}(c), this yields hierarchical patterns,
which reflect the history of the system.
The collapsed zones grow along with the system,
so that earlier-formed zones are larger than more recent ones.

So far, we assumed that the tissue was perfectly homogeneous,
and it may be of interest to examine how disorder,
which is inherent in biological tissues,
would affect the outcome of the model.
As described earlier, disorder can be incorporated as
a multiplicative noise term in the energy of the system.
Instabilities then no longer occur exactly at the locations where the stresses are largest.
Fig.~\ref{images}(d) shows an example of the resulting irregular patterns.
Consistent with theoretical arguments and experimental results concerning crack
patterns~\cite{blumen-93,bohn-05-2},
we find that the effect of disorder depends crucially
on how the distance between collapsed regions compares with the elastic
length scale $\lambda$.
If this distance is much larger than $\lambda$,
then the stresses exhibit a broad plateau
and the location of new collapsed zones is very variable,
being governed essentially by the weak spots due to disorder.
On the other hand,
if the distance between collapsed regions is comparable to or smaller than $\lambda$,
then the stresses have a more localized maximum
and the location of new collapsed zones is close to the location of this maximum.
The latter regime is thus more favorable for obtaining regular and reproducible patterns.
It is worth noting that in growing plant leaves,
the distance between veins is of the same order as the thickness of the leaf~\cite{nelson-97}
(which sets the characteristic length scale for elastic coupling between veins).

\section{Two-dimensional model}

We now turn to the extension of the model to two-dimensional systems.
In this context, the one-dimensional patterns described in the previous section can be interpreted
as parallel stripes of localized deformation.
The more general patterns that we expect to describe
would take the form of networks of stripes of different orientations,
corresponding to different principal directions of deformation.
It is thus clear that a scalar parameter is no longer sufficient to describe the collapsed state.
Since the difference between the two phases lies only in their rest configurations,
we choose here to identify the phase field $\phi$ and $\mathbf\epsilon_r$,
which in 2D is a tensor describing the local rest configuration.
Likewise, the mismatch between the two layers
is now described by a tensor $\mathbf\eta$.
The 2D analog of eq.~\ref{eq_E_eta} thus takes the form
\begin{equation}
E=\int\left[\frac{D}{2}|\nabla\mathbf\epsilon_r|^2
+\frac{\lambda}{2}\mathrm{Tr}(\mathbf\epsilon-\mathbf\epsilon_r)^2+
\mu|\mathbf\epsilon-\mathbf\epsilon_r|^2
+f(\mathbf\epsilon_r)+k(\mathbf u-\mathbf\eta\mathbf x)^2
\right] dS.
\label{2denergy}
\end{equation}
While $\mathbf\eta$ may be anisotropic,
we assume that the tissue has no preferential direction.
Accordingly, $f$ can be expressed as a function of the invariants of $\mathbf\epsilon_r$,
e.g. its trace and its norm.

To obtain a behavior similar to that of the one-dimensional model,
we consider potentials that have a minimum for
$\mathbf\epsilon_r=0$ and a minimum of the same depth for values of $\mathbf\epsilon_r$
corresponding to a smaller natural volume (i.e. $\mathrm{Tr}\mathbf\epsilon_r<0$).
Among all possible potentials satisfying this constraint,
we have chosen the following three illustrative examples.
In the first, the secondary minimum is reached for all deformations
having a given amplitude (norm), regardless of the shear.
The second one specifically favors uniaxial deformations.
In contrast, the third one has a secondary minimum for isotropic compression.

\begin{figure}
\begin{center}
\subfigure[]{
\includegraphics[width=.3\textwidth]{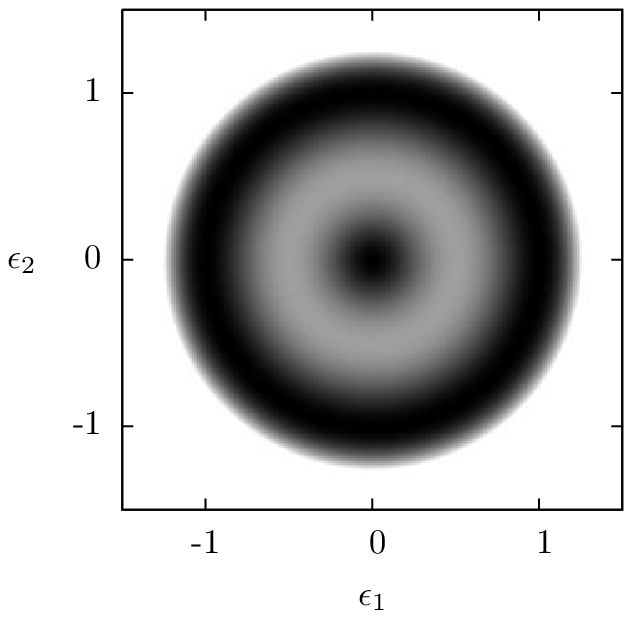}
\label{fig:pot1}
}
\subfigure[]{
\includegraphics[width=.3\textwidth]{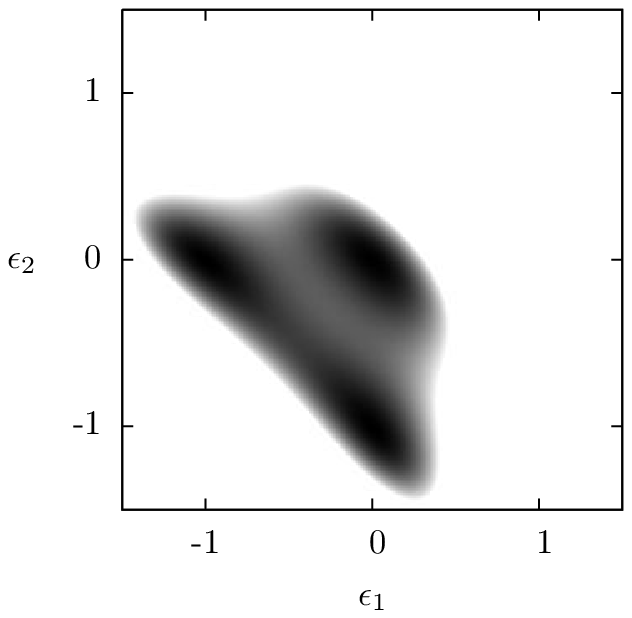}
\label{fig:pot2}
}
\subfigure[]{
\includegraphics[width=.3\textwidth]{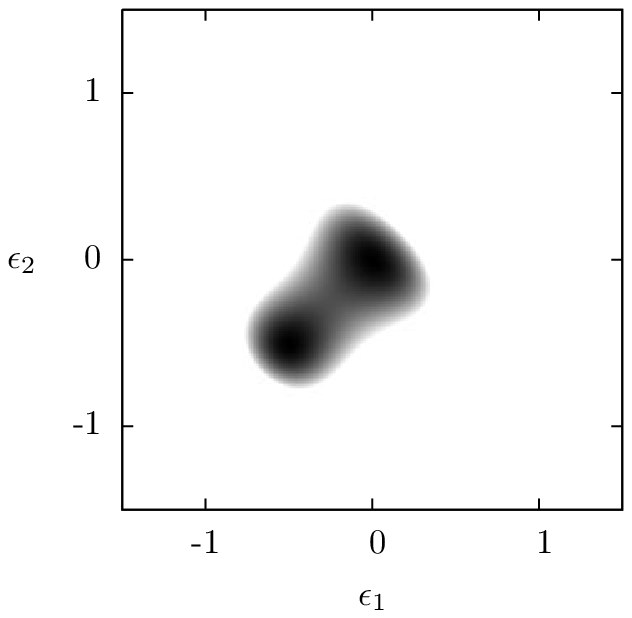}
\label{fig:pot3}
}
\end{center}
\caption{
Gray scale plots of the different potentials $f(\mathbf\epsilon_r)$ used.
$\epsilon_1$ and $\epsilon_2$ are the eigenvalues of $\mathbf\epsilon_r$.
Darker regions correspond to lower values.
(a) The potential defined by eq.~\ref{eq:pot1}
favors compressive strains of a given amplitude,
regardless of the shear.
(b) The potential defined by eq.~\ref{eq:pot2}
and the first set of coefficients given in the text
favors uniaxial compression.
(c) The potential defined by eq.~\ref{eq:pot2}
and the second set of coefficients
favors isotropic compression.
}
\label{fig:pot}
\end{figure}

More specifically, the first potential is defined by
\begin{equation}
\label{eq:pot1}
f(\epsilon_r)=-(|\epsilon_r|-1/2)^2+(|\epsilon_r|-1/2)^4,
\end{equation}
which has minima for $\epsilon_r=0$ and for $|\mathbf\epsilon_r|=1$
(see fig.~\ref{fig:pot1}).
The second and third examples are defined as polynomials of the invariants of $\mathbf\epsilon_r$, i.e.
\begin{equation}
\label{eq:pot2}
f(\mathbf\epsilon_r)=\sum c_{ij}(\mathrm{Tr}\mathbf\epsilon_r)^i|\mathbf\epsilon_r|^{2j}.
\end{equation}
The two sets of coefficients used are
$c_{20}=.4, c_{01}=.2, c_{30}=.7, c_{11}=.5, c_{40}=.4, c_{21}=.1, c_{02}=.1$,
which corresponds to minima for uniaxial deformations
(see fig.~\ref{fig:pot2})
and $c_{20}=.25, c_{01}=.5, c_{30}=1, c_{11}=0, c_{40}=.375, c_{21}=0, c_{02}=.5$,
which leads to a minimum for isotropic compression
(see fig.~\ref{fig:pot3})
\footnote{
Note that the potential defined by eq.~\ref{eq:pot1} cannot be written in this form.
}.

In contrast with the one-dimensional system show in fig.~\ref{system},
in which the edges were free of stresses,
most of the 2D simulations were done with periodic boundary conditions,
which allow the mechanical equilibrium equation
to be efficiently solved in Fourier space.
In this case,
the phase field is initially chosen to be random,
to break the symmetry of the system.
The initial conditions
(mismatch or size for a growing system)
are chosen such than an instability immediately develops,
preventing relaxation to a uniform state.

\begin{figure}
\begin{center}
\subfigure[]{
\includegraphics[width=.3\textwidth]{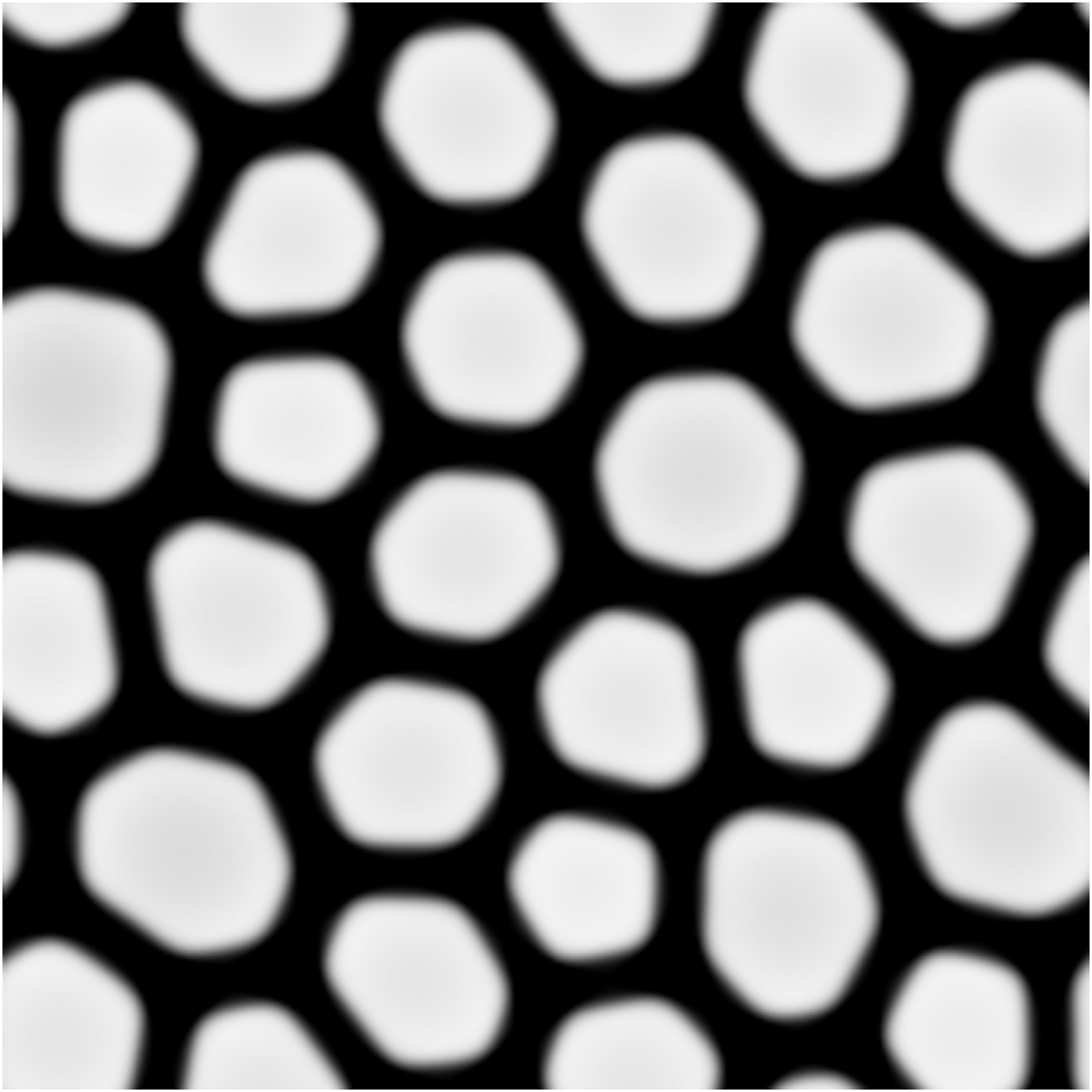}
\label{fig:2drev1}
}
\subfigure[]{
\includegraphics[width=.3\textwidth]{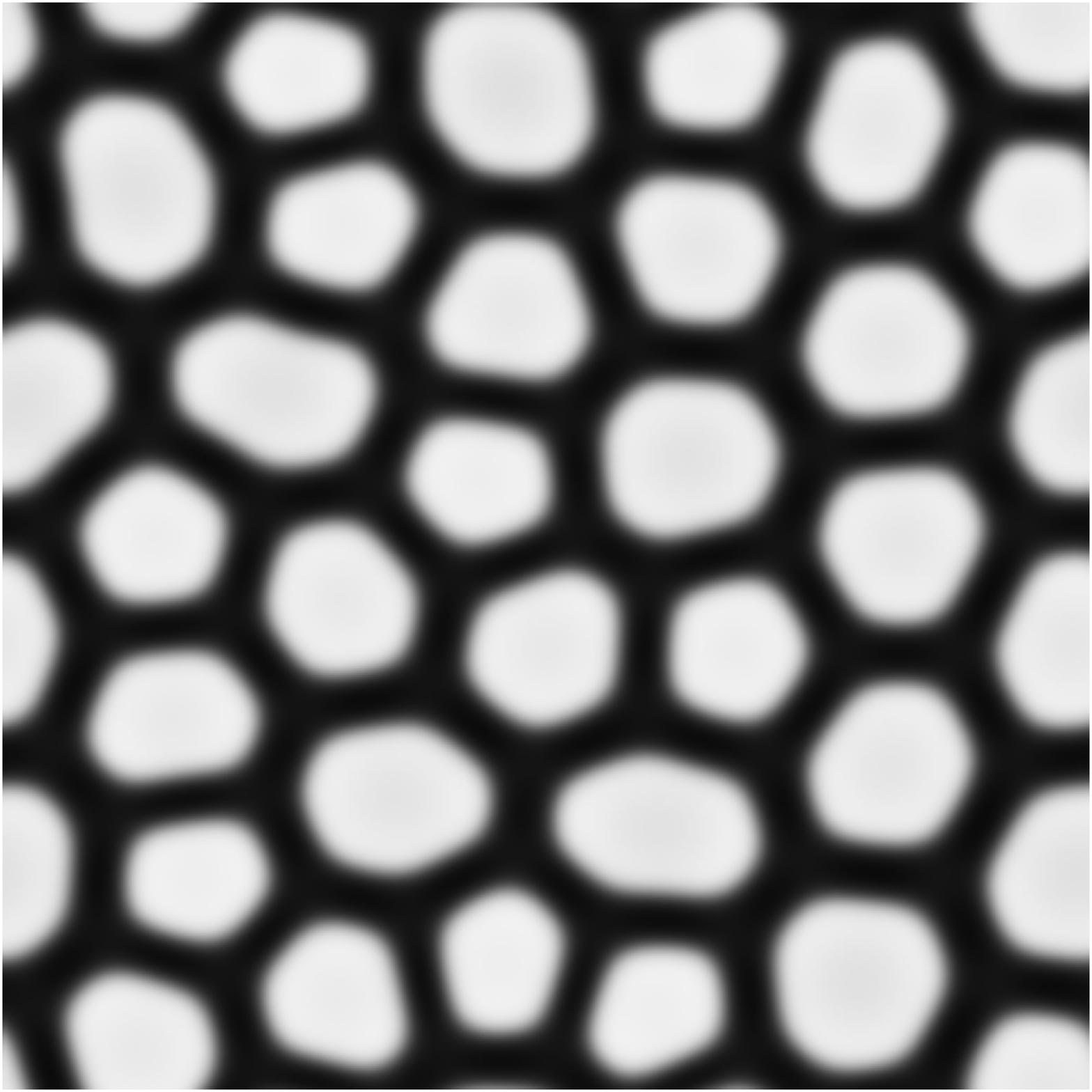}
\label{fig:2drev2}
}
\subfigure[]{
\includegraphics[width=.3\textwidth]{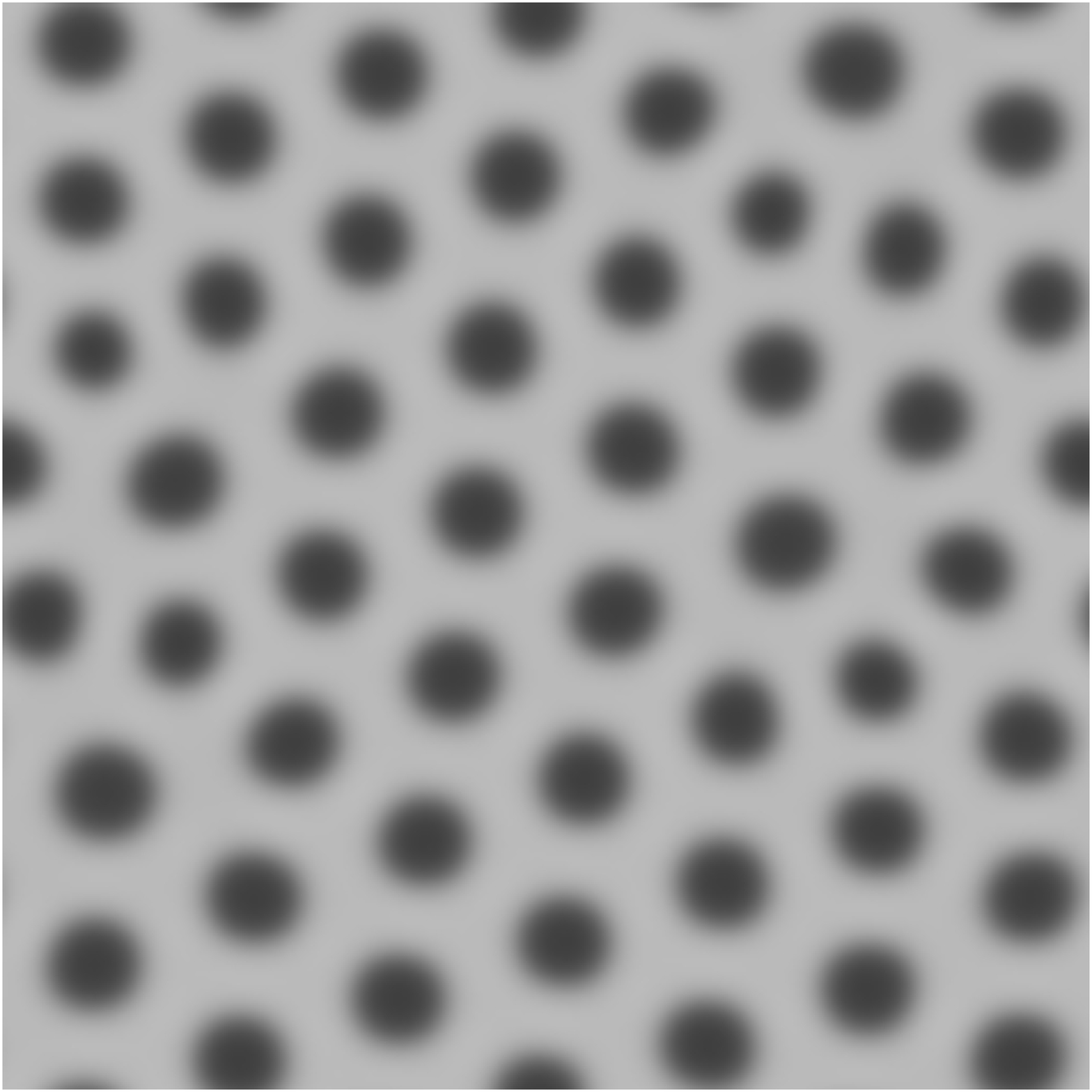}
\label{fig:2drev3}
}
\end{center}
\caption{
Typical patterns observed in a layered two-dimensional system
with the different potentials.
Periodic boundary conditions were used and the initial state was randomly perturbed.
Gray levels correspond to the trace of the rest configuration $\mathbf\epsilon_r$,
with darker areas corresponding to collapsed regions.
Potentials that allow or favor uniaxial compression
(a and b, corresponding to the potentials of figs.~\ref{fig:pot1} and \ref{fig:pot2}, respectively)
yield reticulate patterns,
while the potential that favors isotropic compression
produces discontinuous domains arranged in a hexagonal lattice
(c, corresponding to the potential of fig.~\ref{fig:pot3}).
}
\label{fig:2drev}
\end{figure}

As in the one-dimensional case,
we first consider reversible evolution in a non-growing system (fig.~\ref{fig:2drev}.
Here, the mismatch between the two layers is directly set to its final value,
and the system is allowed to relax to equilibrium.
The first two potentials
(shown in figs.~\ref{fig:pot}(a) and (b)),
which allow or favor uniaxial compression,
yield very similar patterns of interconnected stripes,
which could be likened to crack networks in directional growth systems
such as basalt columns~\cite{Morris2000}.
In contrast,
the potential shown in fig.~\ref{fig:pot3},
which favors isotropic compression,
produces islands of localized deformation arranged in an hexagonal lattice,
which are similar to some structures observed in solid-solid phase
transitions~\cite{lebouar98}.
The characteristic length of the above patterns,
as in 1D, is set by the elastic length scale $\lambda=\sqrt{\mu/k}$.

\begin{figure}
\begin{center}
\subfigure[]{
\includegraphics[width=.4\textwidth]{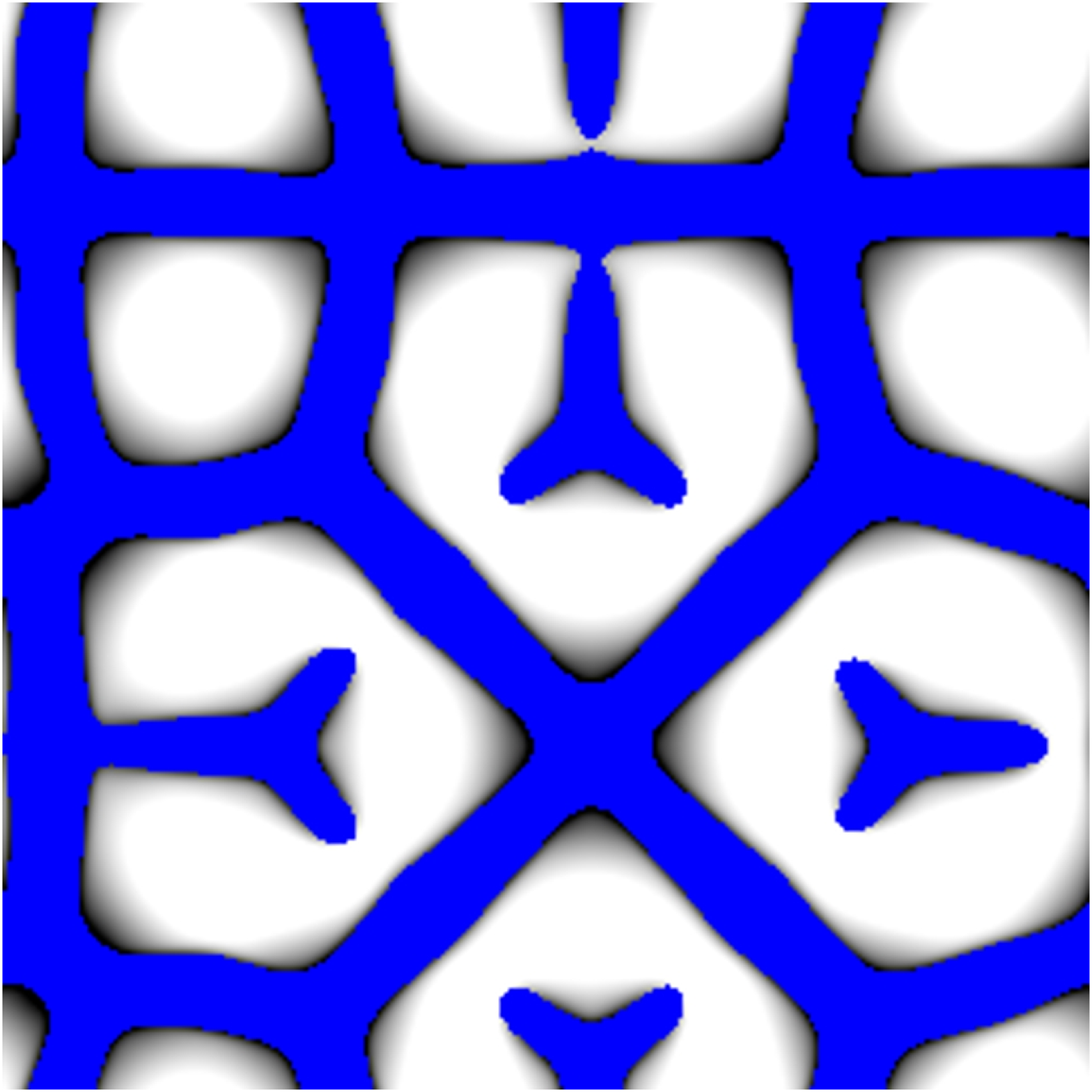}
\label{fig:tensionrev1}
}
\subfigure[]{
\includegraphics[width=.4\textwidth]{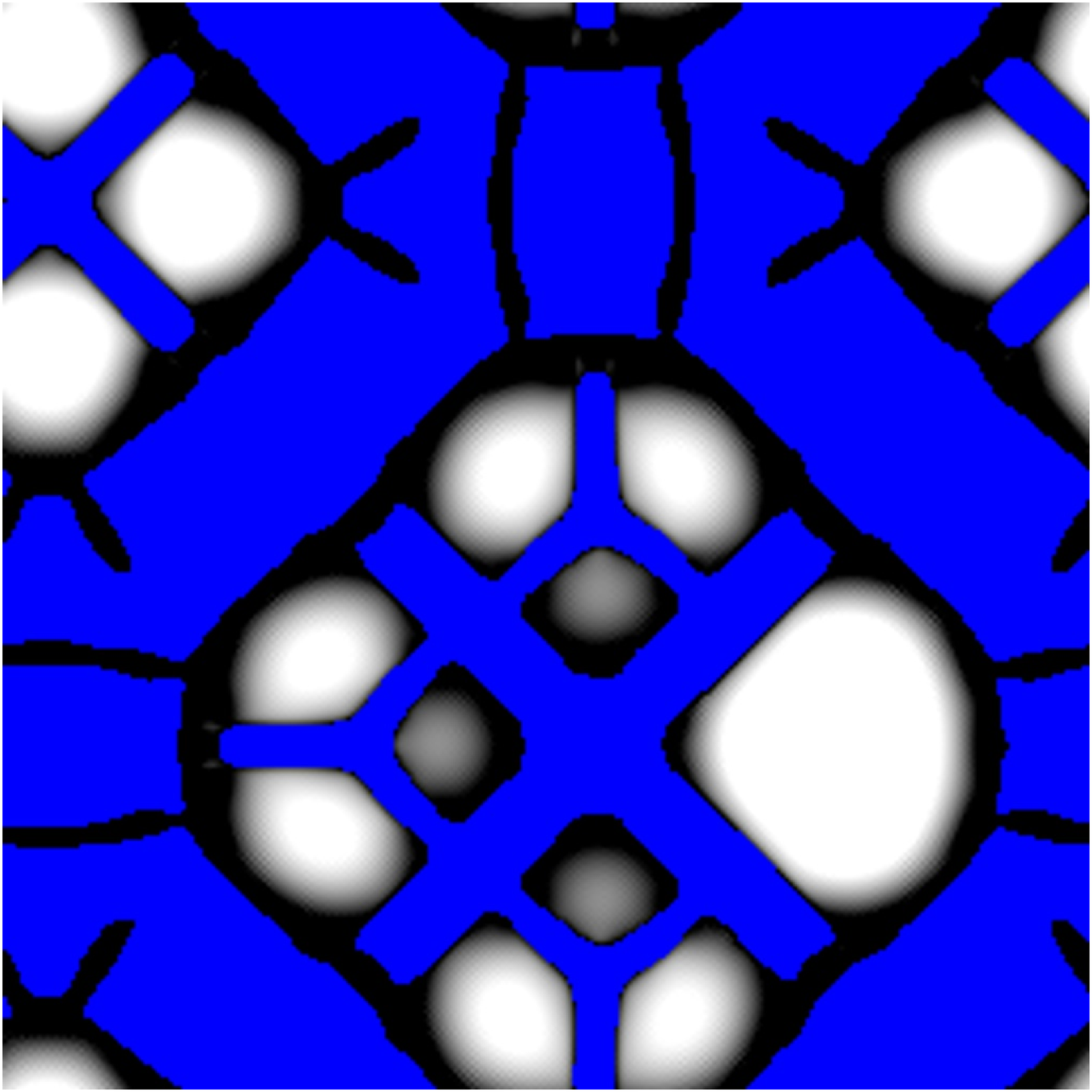}
\label{fig:tensionirrev1}
}
\\
\subfigure[]{
\includegraphics[width=.4\textwidth]{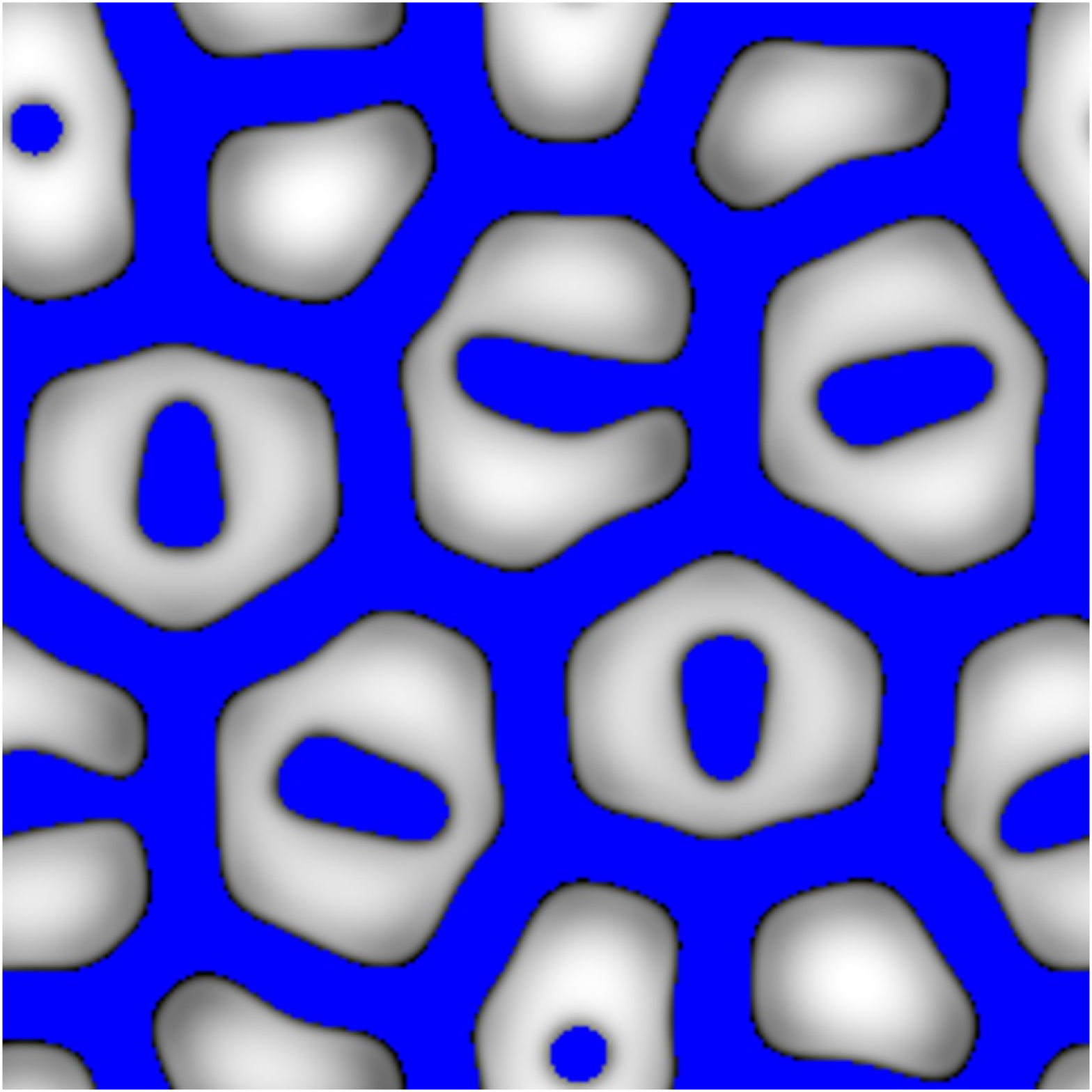}
\label{fig:tensionrev2}
}
\subfigure[]{
\includegraphics[width=.4\textwidth]{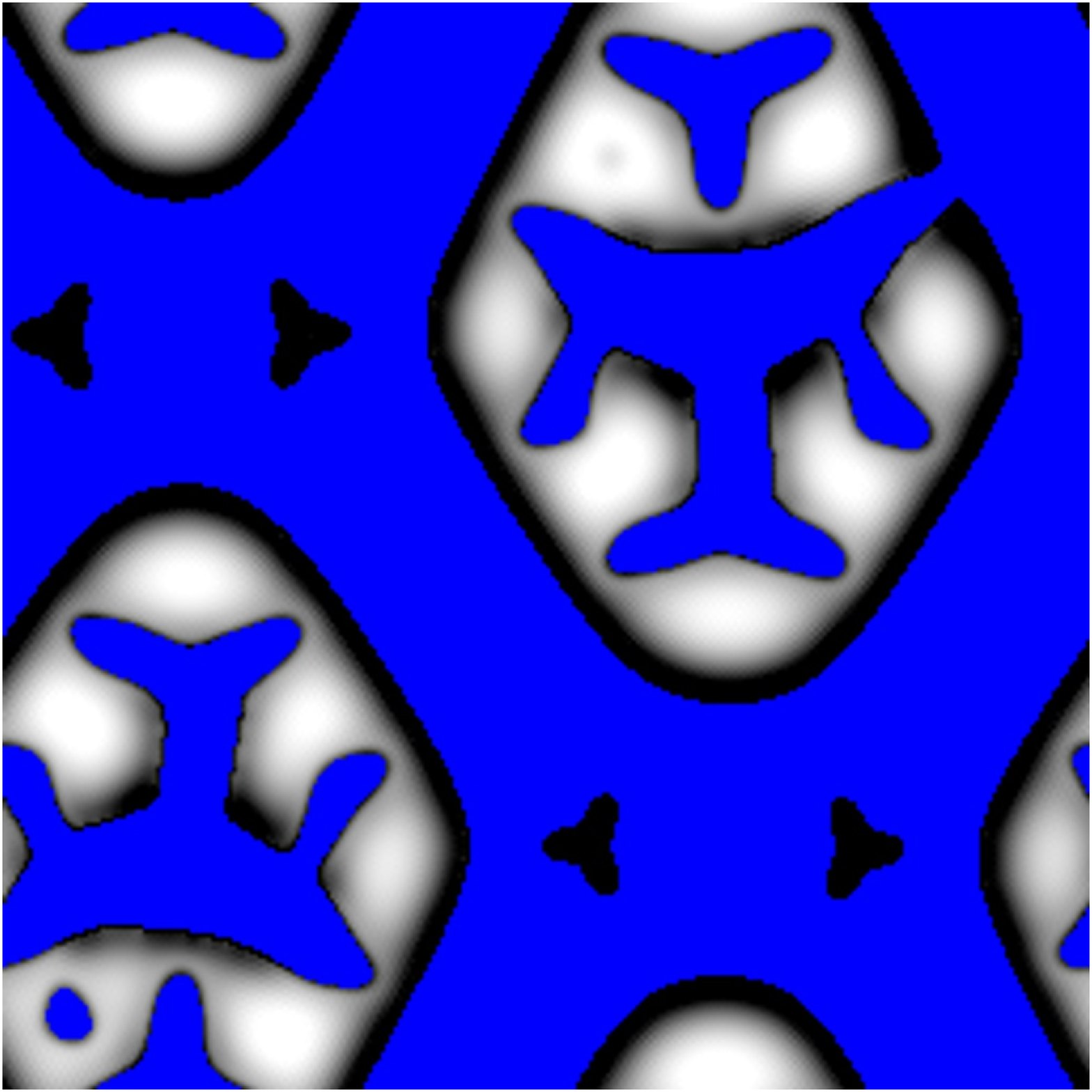}
\label{fig:tensionirrev2}
}
\end{center}
\caption{
(Color online)
Effect of irreversibility on the mechanical state of the system and pattern formation.
All figures represent growing systems with periodic boundary conditions.
(a) and (b) were obtained with the potential of fig.~\ref{fig:pot1},
(c) and (d) with the potential of fig.~\ref{fig:pot2}.
Evolution was reversible in (a) and (c), irreversible in (b) and (d).
Collapsed regions (defined by thresholding the value of $\mathrm{Tr}(\mathbf\epsilon_r)$) are shown in blue.
Gray levels indicate compressive stresses
(lighter regions are more strongly compressed),
and regions appearing in black are under tension.
In the reversible case,
the stresses decrease near the collapsed regions,
yet generally remain compressive.
In the irreversible case,
this is also true for recently formed structures.
In contrast,
the larger, older structures generate tensile stresses.
These tensile stresses hinder the progression of newly formed structures,
yielding disconnected patterns.
}
\label{fig_residual_stress}
\end{figure}

The difference between the two potentials
that produce reticulate patterns
becomes more apparent when considering a growing system (and a reversible
evolution),
which makes it possible to observe individual nucleation events leading to the
formation of new collapsed regions.
Indeed,
in the case of the potential of fig.~\ref{fig:pot1},
new collapsed regions take the form of trijunctions
that split an uncollapsed region into three\footnote{A typical example in the
case of an irreversible evolution can be seen in fig.~\ref{fig_residual_stress}(a)},
while in the case of the potential of fig.~\ref{fig:pot2},
they occur as single stripes that divide an uncollapsed region
across its longest extension \footnote{A typical example can be seen in the case
of an irreversible evolution can be seen in fig.~\ref{fig_residual_stress}(c)}.
This qualitative difference can be explained by the fact
that the core of a trijunction is compressed in both directions,
a state that is disfavored by the potential of fig.~\ref{fig:pot2}.

Finally, we consider irreversible evolution in a growing system,
which we found to lead to hierarchical patterns in 1D. 
In that case,
irreversibility was implemented by forcing the evolution of the scalar
phase field to be monotonic once a certain threshold was exceeded.
A straightforward extension to two dimensions
would be to apply the same rule to the trace of the rest configuration tensor $\mathbf\epsilon_r$.
However,
this would allow the principal directions of the rest configuration in the collapsed phase to vary
(and simulations indicate that this can indeed occur).
Instead,
we wish the establishment of these special directions to be definitive,
as is the transition to the collapsed state
(think of the elongation of vascular cells in plant leaves).
Accordingly,
we impose that, once the compression in some direction
has exceeded a certain threshold, it can no longer decrease
\footnote{
To this end, we define a ``maximum deformation tensor'' $\mathbf\epsilon_m$
that keeps track of the maximum past compression in any direction,
i.e. the tensor $\mathbf\epsilon_r(t)-\mathbf\epsilon_m(t')$
is positive for all times $t<t'$.
From this maximum deformation tensor,
we define an ``irreversible deformation tensor'' $\mathbf\epsilon_i$.
In a basis that diagonalizes $\mathbf\epsilon_m$,
$\mathbf\epsilon_i=\mathrm{diag}(f(\lambda_1),f(\lambda_2))$,
where $\mathbf\epsilon_m=\mathrm{diag}(\lambda_1,\lambda_2)$.
The function $f$ is defined by $f(x)=0.9x$ if $x<-0.5$ and $f(x)=\infty$ if $x>0.5$,
so that compressions in excess of $.5$ are stored in $\epsilon_i$.
Irreversibility is implemented by imposing that
the tensor $\mathbf\epsilon_r-\mathbf\epsilon_i$ is negative.
The factor $.9$ in the definition of $f$ was introduced to prevent numerical instability.
}.

\begin{figure}
\begin{center}
\begin{tabular}{ccc}
\begin{tabular}[c]{c}
\includegraphics[width=.35\textwidth]{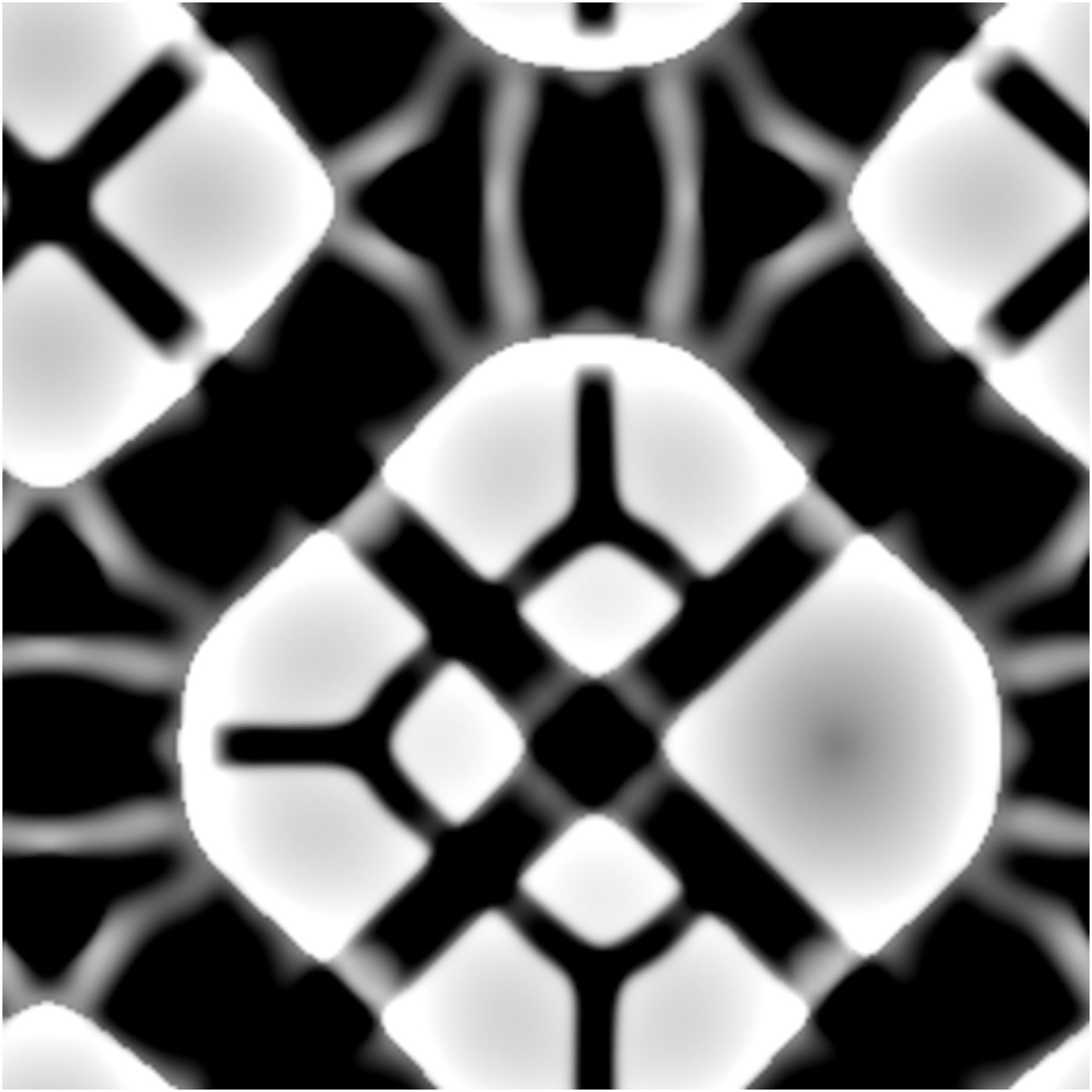}
\\
{\small(a)}
\end{tabular}
&
\begin{tabular}[c]{c}
\includegraphics[width=.35\textwidth]{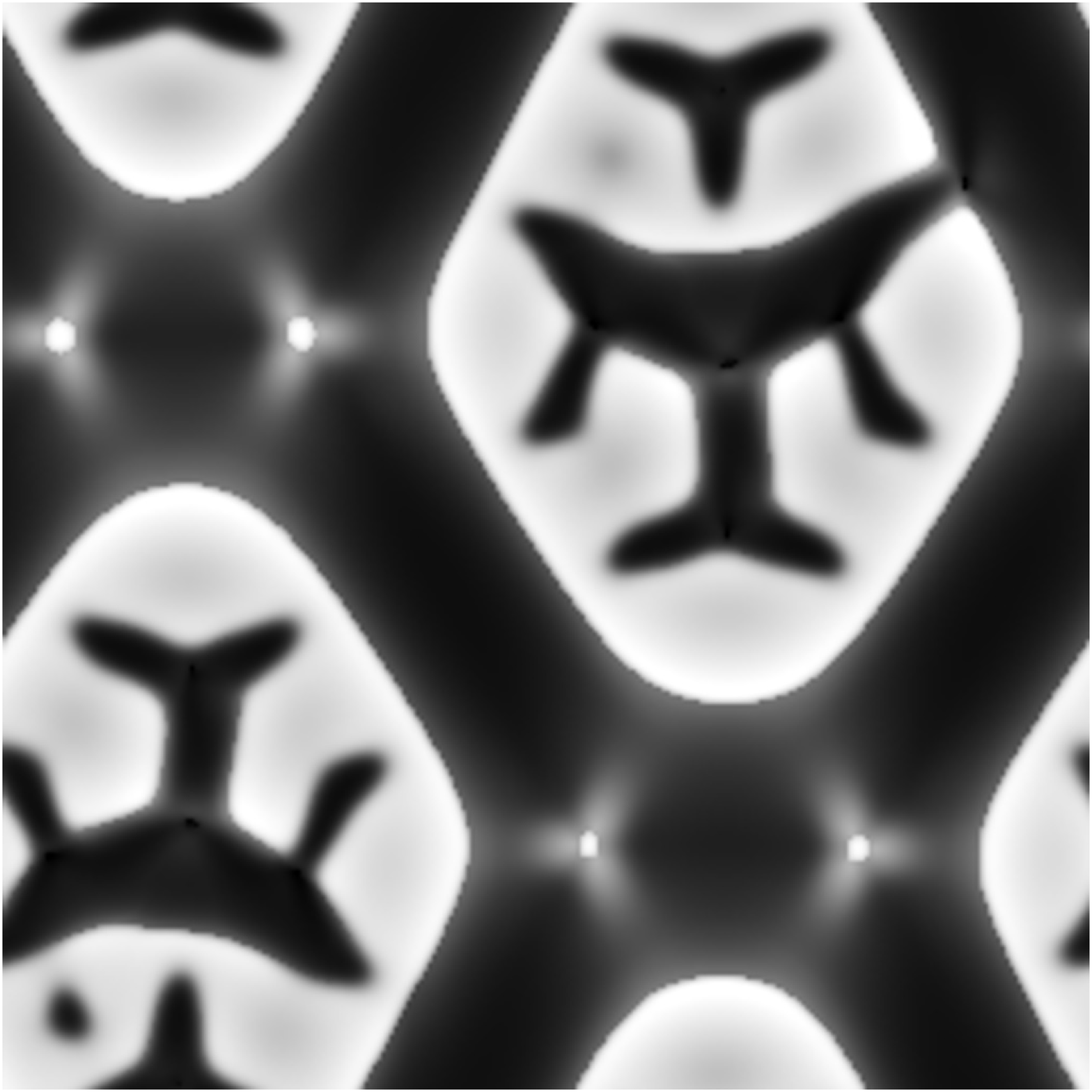}
\\
{\small(b)}
\end{tabular}
&
\begin{tabular}[c]{c}
\includegraphics[width=.14\textwidth]{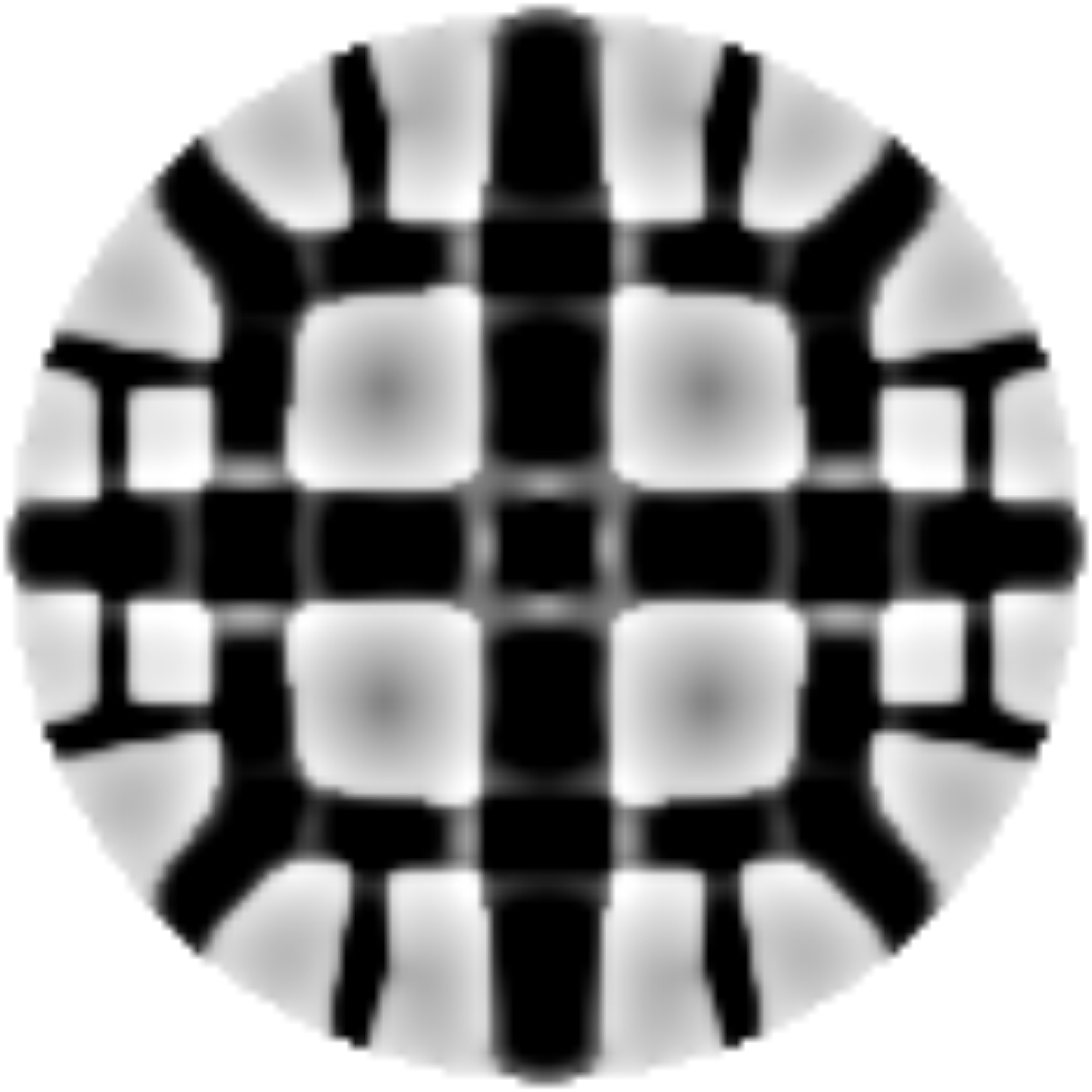}
\\
{\small(c)}
\\
\includegraphics[width=.14\textwidth]{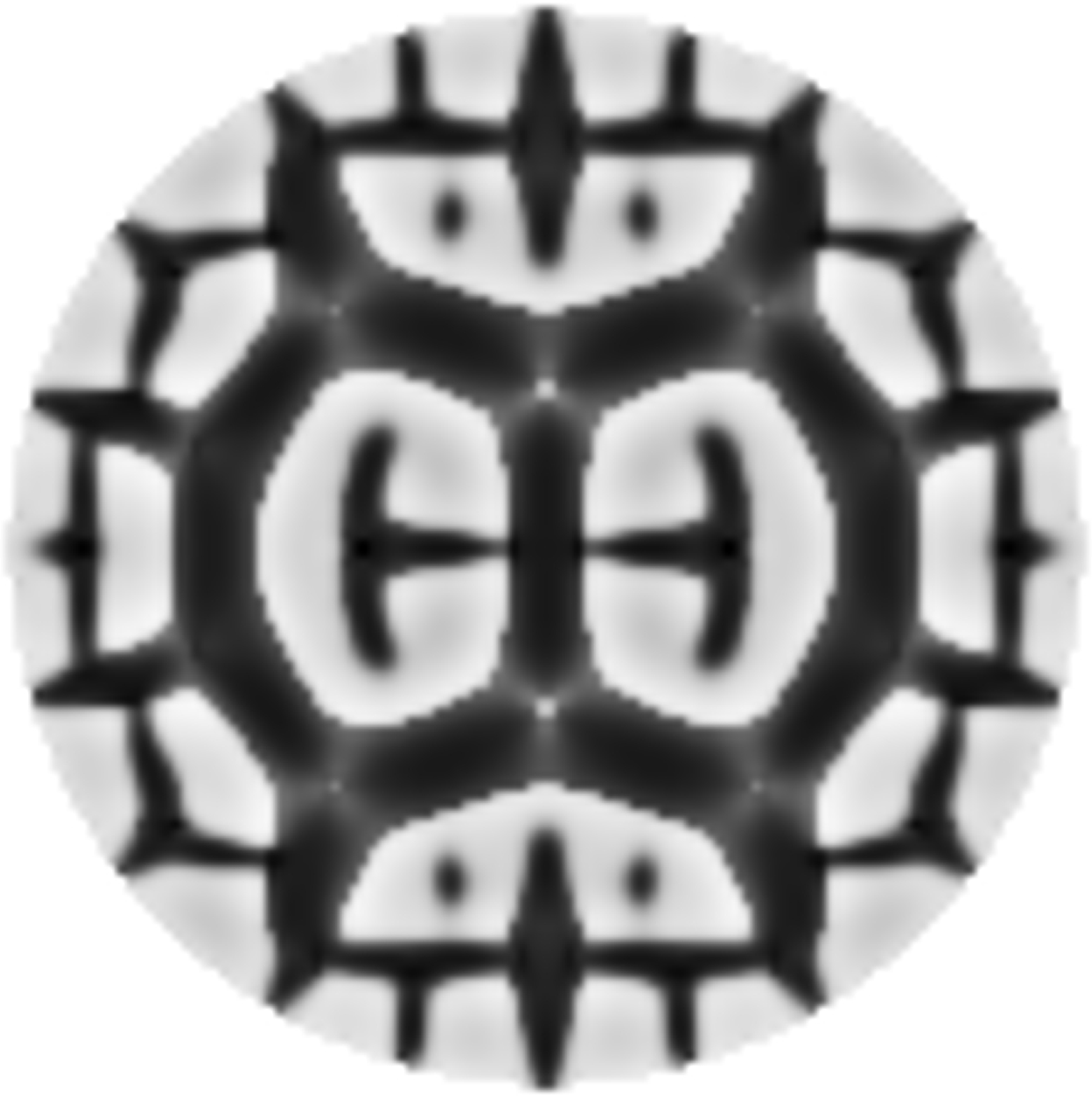}
\\
{\small(d)}
\end{tabular}
\end{tabular}
\end{center}
\caption{
Patterns generated by irreversible evolution in growing systems
with different potentials and boundary conditions
(periodic boundary conditions or circular domains).
(a) and (c) were obtained with the potential of fig.~\ref{fig:pot1},
(b) and (d) with the potential of fig.~\ref{fig:pot2}.
As expected, hierarchical structures are formed.
However, they are disrupted by the tensions
that accumulate within and around the collapsed regions as they grow
(see text and fig.~\ref{fig_residual_stress}).
Notice the lighter spots or stripes within the collapsed regions.
}
\label{fig_2d_hierarchical}
\end{figure}

As in 1D, we consider a growing system with a fixed mismatch,
and find that hierarchical structures are obtained.
As shown on fig.~\ref{fig_2d_hierarchical},
the shape of the patterns depends both on the potential used and on the
shape of the growing domain (for the earliest structures that are formed).
However,
we find that the simple model presented here produces unexpected artifacts,
such as regions of lower compression within the collapsed zones,
and open-ended collapsed regions that fail to reconnect with
previously formed structures.

As illustrated by fig.~\ref{fig_residual_stress},
these effects are caused by tensions that
accumulate in the collapsed structures as they grow.
When the transition to the collapsed state is reversible,
the collapsed regions tend to an equilibrium width
such that compressive stresses are relaxed in their neighborhood
(i.e. for an ideal, infinitely long and straight stripe,
the stress at the interface has a fixed value
equivalent to the plateau stress $\sigma_p$ of the one-dimensional model).
In contrast,
in the irreversible case,
the width of each collapsed region grows along with the system
and exceeds its equilibrium width.
Instead of just releasing the compression in their vicinity,
these growing collapsed regions give rise to tensile stresses.
As can be seen on fig.~\ref{fig_2d_hierarchical},
these tensile stresses both destabilize the collapsed regions,
and prevent the progression of new collapsed regions towards them.
It must be noted that such tensile stresses are also generated in the 1D model.
However, they do not have such dramatic effects in that case,
because new collapsed structures always form away from existing ones.

\section{Discussion}
In this article,
we explored a possible mechanism for biological patterning by mechanical stresses.
This mechanism,
which involves the stress-mediated transition of a tissue between two states,
was shown to yield a response similar to that of a non-linear elastic material,
justifying an analogy with mechanical instabilities.
The formation of patterns was driven by coupling the tissue to a rigid substrate,
which could represent a stiffer tissue such as the epidermis of plant leaves.
As in the mechanical instabilities of layered physical systems,
such as the wrinkling of a film bound to a substrate,
this elastic coupling introduces a characteristic length scale
that governs the size of the patterns.

We first analyzed a one-dimensional system.
In this case, the model involves a small number of parameters,
and it is possible to describe the equilibrium patterns analytically.
Numerical simulations showed that regular patterns,
which are largely independent of the history of the system,
are obtained when the transition between the two states of the tissue is reversible.
In contrast,
when this transition is made irreversible to represent tissue differentiation,
the history of the system is retained in its final state.
In a growing system, new structures keep forming over time,
yielding hierarchical patterns.
This is very similar to the development of leaf venation networks,
in which veins of different orders form successively
over the course of leaf growth~\cite{nelson-97}.

In two dimensions, because of the tensorial nature of elastic fields,
the potential that characterizes the preferred states of the tissue
can have many different forms.
While we did not carry out a systematic analysis,
the different examples considered
suggest that reticulate patterns can readily be obtained when the potential
allows or favors uniaxial deformations of the tissue.
In contrast, islands of deformed tissue
are obtained with a potential that favors isotropic deformations.
This suggests that a mechanical model of leaf venation patterning
could account for both wild-type patterns
and the disconnected vasculature observed in certain mutants~\cite{koizumi-00}.

A very similar model of leaf vein patterning
was recently proposed in~\cite{jagla}.
In that approach, however, the different states of the tissue
are characterized by a single scalar order parameter.
The transition to the collapsed state is associated with an isotropic deformation,
and the collapsed tissue is assumed to have a reduced
shear modulus to allow the formation of reticulate patterns.
The elongation of vascular cells is described as
a change of their rest volume combined with a strong elastic shear. 
One benefit of this approach is that it involves
a much smaller number of parameters.
However,
it is unclear that it gives a realistic description of the mechanical state of vascular cells.
While it comes at the cost of a 	greater complexity,
we find it more appropriate to describe the elongation of vascular cells
as a change in their rest shapes.
This also makes it possible to address the effect of variations
in these rest shapes on the overall pattern.
For instance, as mentioned above,
our results suggests that failure to elongate properly
can lead to disconnected patterns
such as are observed in some mutants.

From the preceding results,
we expected that the introduction of irreversibility in a growing
two-dimensional system would lead to hierarchical, reticulate patterns.
However, we found that strong tensile stresses develop
in collapsed regions as they grow beyond their initial, equilibrium size,
and dramatically affect the evolution of the system.
This is a consequence of the very simple growth laws assumed in our model.
The rest configuration of the tissue changes when it switches between its two states,
but growth in each of the two states is uniform and proceeds at a constant rate,
allowing residual stresses to accumulate.
In reality, the growth of biological tissues
depends on the mechanical stresses to which they are subjected~\cite{barley1962,cowin-04},
which can be expected to limit the buildup of residual stresses.
Incorporating this dependence would be essential
to further investigate biological patterning by mechanical forces.
Another perspective would be to integrate  the mechanical approach developed
here with the biochemical factors of tissue differentiation.
For instance, the differentiation of vascular cells in plant leaves involves
the expression and polar localization of auxin carrier proteins~\cite{scarpella-06},
and it would be of interest to investigate how these processes
are connected with the elongation of vascular cells.

\section*{Acknowledgements}

This work was supported by EC NEST project MechPlant. The authors would like to
thank one anonymous referee for detailed remarks and helpful suggestion on a
previous version of this manuscript. 

\end{document}